\documentclass[useAMS,usenatbib]{mn2e}
\usepackage{epsfig}

\usepackage{amsmath}
\usepackage{amssymb}

\title[Magnetic Braking of Ap/Bp Stars]{Magnetic Braking of Ap/Bp 
Stars: Application to Compact Black-Hole X-Ray Binaries}
\author[Justham, Rappaport and Podsiadlowski]{Stephen 
Justham,$^{1}$\thanks{Email: sjustham@astro.ox.ac.uk} Saul 
Rappaport$^{2}$ and Philipp Podsiadlowski$^{1}$ \\
$^{1}$ Department of Astrophysics, Oxford University, OX1 3RH \\
$^{2}$ Department of Physics, MIT, Cambridge, Massachusetts, MA 02139, USA}

\date{Submitted .... Accepted ....}

\pagerange{\pageref{firstpage}--\pageref{lastpage}}
\pubyear{2005}

\begin{document}
\maketitle

\label{firstpage}

\begin{abstract}
We examine the proposal that the subset of neutron-star and black-hole
X-ray binaries that form with Ap or Bp star companions will experience
systemic angular-momentum losses due to magnetic braking, not otherwise
operative with intermediate-mass companion stars.   We suggest that for
donor stars possessing  the anomalously high magnetic fields associated
with Ap and Bp stars, a magnetically coupled, irradiation-driven stellar
wind can lead to substantial systemic loss of angular-momentum. 
Hence these systems, which would otherwise not be expected to experience
`magnetic braking', evolve to shorter orbital periods during mass transfer. 
In this paper we detail how such a magnetic braking
scenario operates.  We apply it to a specific astrophysics problem
involving the formation of compact black-hole binaries with low-mass
donor stars.  At present, it is not understood how these systems form,
given that low-mass companion stars are not likely to provide sufficient
gravitational potential to unbind the envelope of the massive progenitor
of the black hole during a prior ``common-envelope'' phase. 
On the other hand, intermediate-mass companions, such as Ap and Bp stars,
could more readily eject the common envelope.  However, in the absence
of magnetic braking, such systems tend to evolve to long orbital periods. 
We show that, with the proposed magnetic braking properties afforded by Ap
and Bp companions, such a scenario can lead to the formation of compact
black-hole binaries with orbital periods, donor masses, lifetimes,
and production rates that are in accord with the observations. 
In spite of these successes, our models reveal a significant discrepancy
between the calculated effective temperatures and the observed spectral
types of the donor stars.  Finally, we show that this temperature 
discrepancy would still exist for other scenarios invoking initially 
intermediate-mass donor stars, and this presents a substantial 
unresolved mystery.
\end{abstract}


\section[]{Introduction}
Magnetic braking is a relatively poorly understood mechanism for angular
momentum loss in single stars as well as compact binaries, but is often
invoked as the driver of mass transfer in compact binaries such as
cataclysmic variables and low-mass X-ray binaries
\citep[see, e.g.,][]{VZ81, RVJ83, SR83}.  Models of magnetic
braking usually invoke two key components: a large-scale
magnetic field and a stellar wind. When coupled together, the field provides
a long lever-arm which makes the wind an efficient angular-momentum sink.

According to conventional thinking, braking via a magnetically constrained
stellar wind is inoperative in stars with radiative envelopes, 
\citep[see, e.g.][]{K88} and
thus is confined to stars with $M \lesssim 1.5~M_\odot$.  However, a
subclass of intermediate-mass stars (the Ap and Bp stars) possess
substantial magnetic fields; herein we suggest that an
irradiation-driven stellar wind combined with their magnetic fields
would lead to significant magnetic braking in compact binaries 
containing such stars.

We apply this heretofore unutilized magnetic braking to the formation of
compact black-hole X-ray binaries.  These systems are not easy to 
account for 
within previous formation scenarios -- yet they comprise 
about half of the well-studied 
black-hole X-ray binaries. Nine of 
the seventeen systems listed in \citet{LBW}, 
\citet{RKcat} and 
\citet{PRH03} (hereinafter PRH, and references therein) have
orbital periods $< 1$ d and inferred donor masses $\lesssim 1~{\rm{M}}_{\odot}$
\citep[see also][]{McCR}. The empirical estimates of \citet{W96} and
\citet{Rom98} suggest that the Galactic population of short-period
black-hole binaries may number over 1000 systems.

The standard formation scenario for such compact black-hole binaries 
typically 
requires a low-mass donor to have ejected the envelope of 
its primordial binary
companion -- the massive black-hole progenitor -- as it spirals 
inwards during a common-envelope phase 
\citep[see, e.g.][also PRH and references therein]{dKvdHP}. 
The orbital period is reduced from years to days, and 
the core of the massive star continues to evolve toward collapse and 
the formation of the black hole in the system. However, it is not 
clear how the energy required to unbind the
envelope of the black-hole progenitor can be provided by a secondary
star with a mass of only $\lesssim 1.5~{\rm{M}}_{\odot}$
\citep[e.g.][PRH; see the Appendix for a brief summary of this argument]{PCR95, PzVE97, K99}. 
In the conventional scenario,
this problem cannot simply be resolved by assuming a more massive
secondary.  Donor stars with $M \gtrsim 1.5~M_\odot$ are not expected
to evolve into the short-period population, as they are thought {\em
not} to be subject to canonical magnetic braking; mass transfer from 
the less massive to the more massive component of the system tends 
to widen the orbit (a consequence of conservation of angular-momentum).

In addition to this energetics problem, standard assumptions about star
formation also suggest a demographic difficulty -- even if the envelope
of the massive primary could be ejected by the spiral-in of a low-mass
star, there are expected to be relatively few primordial binaries with
such an extreme mass ratio as required in that standard picture 
\citep[e.g.][]{GCM80, FSS05}.

Hence, attempts to produce these low-mass black-hole binary systems have
sometimes been exotic, postulating -- for example -- descent from a
triple system \citep{EV86}, that the companion and black hole both
form from a Thorne--${\rm{\dot{Z}}}$ytkow object \citep{PCR95}, or as
a result of runway accretion in the merger of a neutron-star with a
massive star (PRH). Such ideas have been considered since the `simple'
alternatives often have far-reaching consequences, e.g., modifying the
energetics of common-envelope evolution by a large factor to allow a
low-mass star to eject the envelope of the primary is unpalatable,
whilst magnetic braking from intermediate-mass radiative stars
is not otherwise expected.

For standard input physics, PRH concluded that ``black-hole binaries
with low-mass secondaries can only form with apparently unrealistic
assumptions'', which is an unsatisfactory position for binary
evolutionary theory.  PRH demonstrated that for the expected range of
common-envelope parameters, black-hole binaries are much more likely
to form with intermediate-mass companions than low-mass companions.
They also pointed out that to form the observed short-period
population from these intermediate-mass systems, an additional
angular-momentum loss mechanism, such as anomalous magnetic braking,
would be required.

An elegant way to solve this problem of forming short-period 
black-hole binaries would be to find such an angular-momentum
loss mechanism for intermediate-mass stars. Ideally it would operate
only on a subset of the intermediate-mass donor population,
such that the long-period and short-period black hole binary populations
can both be formed. This paper suggests a self-consistent and plausible
way for initially intermediate-mass companions to lead
to the low-mass companions currently observed, at {\emph{both}} long
and short orbital periods.

Though the bulk of intermediate-mass stars are not expected to undergo
magnetic braking, a fraction of such stars are known to have
anomalously strong magnetic fields, the so-called Ap and Bp stars
\citep{M89, BS04} with surface field strengths ranging from 100 G to
over 10,000 G.  When such a magnetic field is combined with an
irradiation-driven wind from the surface of the donor star, a
plausible angular-momentum loss mechanism is obtained. This paper is
motivated by the realisation that such a mechanism can act on a
portion of the intermediate-mass progenitor population to produce the
short-period low-mass binary companions to black holes. Only a small
portion of the intermediate-mass donors are affected -- those with
anomalously high field strengths. As we will show in Section~3, the
lifetimes of systems with high magnetic fields are significantly
longer, since the initial high mass-transfer rate quickly reduces the
mass of the donor stars, making them evolve as lower-mass stars with
much longer evolutionary timescales. This increases the relative
fraction of the short-period black-hole binary population.

We find that the proposed magnetic braking mechanism should function 
as anticipated, driving systems to short orbital periods and 
producing low-mass donors.  However, we also find that low-mass 
donor stars which descended from intermediate-mass progenitors have 
rather higher temperatures than can be 
consistent with the spectral 
types observed for the donor stars in short-period black-hole X-ray 
binaries. If no reconciliation of this discrepancy is found, this 
result is itself potentially very illuminating, as it would apply to 
any formation channel which invokes primordially intermediate-mass 
donor stars. This would imply that either our understanding of 
common-envelope ejection needs
severe revision, or an ``exotic'' channel is responsible for the
formation of short-period black-hole binaries.

In section 2 we assemble the relevant physics to quantify the 
proposed anomalous magnetic braking mechanism and investigate its 
properties. In Section 3 we demonstrate how this magnetic braking 
naturally produces 
the observed short-period black-hole binaries.  Finally, we consider
observational tests of the model and illustrate its successes and 
major shortcoming.

\section[]{Assumptions and Derivations}

Magnetic braking removes rotational angular-momentum from a star when
its stellar wind remains coupled to the stellar magnetic field as it
leaves the star \citep[e.g.][]{WD67, MS87, K88}. We make the standard
assumption that the wind corotates out to the magnetospheric radius,
defined as the radius where the magnetic pressure and ram pressure 
are balanced,
i.e.,
\begin{equation}
\frac{1}{2} \rho v^{2} \simeq \frac{B^{2}}{8 \pi} = 
\frac{{B_{\rm{s}}}^{2}} {8 \pi} \frac{R_d^6} {r^{6}}~~~,
\end{equation}
where $r$ is the radial distance from the centre of the star, $\rho$
and $v$ are the density and velocity of the stellar wind, respectively,
$B_s$ is the magnetic field strength at the surface of the star, and
$R_d$ is the radius of the donor star.  Here we have assumed a dipolar
magnetic field structure. We can then combine eq. (1) with the continuity
of mass applied to the stellar wind-loss rate, $\dot{M}_{\rm{wind}}$:
\begin{equation}
\dot{M}_{\rm{wind}} = \rho v 4 \pi r^2 ~~~,
\end{equation}
and the assumption that the wind velocity at the surface of the star 
is of order the escape speed:
\begin{equation}
v \simeq \sqrt{ \frac{2 G M_{\rm{d}}} {R_d} } ~~~
\end{equation}
(where $M_d$ is the mass of the donor star under consideration) to 
yield the magnetospheric radius:

\begin{equation}
\label{MagRad}
r_{\rm{m}} \simeq {B_s}^{1/2} {R_d}^{13/8} 
{\dot{M}_{\rm{wind}}}^{-1/4} {(GM_{\rm{d}})^{-1/8}} ~~~.
\end{equation}
We note that assigning a velocity to the wind by assuming it is equal to
the escape speed at the {\em magnetosphere}, rather than the surface
of the star, recovers the expression due to \cite{LPP73}. We have
performed the derivations below using both versions of the
magnetospheric radius, and the differences in outcome are minimal.

The rate of change of angular-momentum due to magnetic braking, 
$\dot{J}_{\rm{MB}}$, is then:
\begin{equation}
\label{MBtorque}
\dot{J}_{\rm{MB}}=-\Omega_{d} {r_{\rm{m}}}^{2} 
\dot{M}_{\rm{wind}}=-\Omega_d B_{\rm s}{R_d}^{13/4} 
{\dot{M}_{\rm{wind}}}^{1/2} {(GM_{\rm{d}})^{-1/4}} ~,
\end{equation}
where $\Omega_{d}$ is the angular rotation frequency of the donor 
star, and we have substituted in the expression for $r_m$ from eq. 
(4). We expect that $\Omega_{d} = \Omega_{\rm orb}$ for close 
binaries as they are likely to be tidally locked.

\subsection{Estimate of the Required Wind-Loss Rate}

Before proceeding, we first estimate what angular-momentum loss rate,
$\dot J_{\rm MB}$, will prevent the mass transfer in the binary (from
the lower-mass donor to the higher-mass black hole) from widening the orbit
and, in fact, will allow for orbital {\em shrinkage}.  To make a rough
estimate of the effect of $\dot J_{\rm MB}$ on the orbital separation,
we assume conservative mass transfer and, from angular-momentum
conservation, obtain:
\begin{equation}
\frac{\dot J_{\rm MB}}{J}= \frac{\dot a}{a}+\frac{\dot M_{\rm 
RLOF}}{M_d}\left(1-\frac{M_d}{M_{\rm BH}}\right) ~~~,
\end{equation}
where $\dot M_{\rm RLOF}$ is the mass-loss rate from the donor star
that is transferred to the black hole via Roche-lobe overflow, and $a$ is the
orbital separation.  If we require that $\dot a < 0$, and solve for
$\dot J_{\rm MB}$, neglecting the mass of the donor star compared to
that of the black hole, we find:
\begin{equation}
\dot J_{\rm MB} \gtrsim \dot M_{\rm RLOF} \left(G M_{\rm 
tot}a\right)^{1/2} ~~~,
\end{equation}
where $M_{\rm tot}$ is the total mass of the binary.  When we combine
this expression with the middle term in eq. (5) and assume that the
donor star is corotating with the orbit, we find:
\begin{equation}
\dot{M}_{\rm wind} ~ r_m^2  \gtrsim   \dot{M}_{\rm RLOF} ~ a^2 ~~~.
\end{equation}

If we now parameterise the orbital separation in terms of the donor
star radius and the dimensionless Roche-lobe radius, $r_L$
\citep[e.g.][]{PPE83} as $a= R_s/r_L$, and use the magnetospheric radius from
eq. (\ref{MagRad}) we find the requirement for $\dot M_{\rm wind}$:
\begin{equation}
\dot{M}_{\rm{wind}} \gtrsim \frac{\sqrt{G M_{\rm d}}} {B_s^2 
R_d^{5/2}r_L^4} {\dot{M}_{\rm{RLOF}}}^{2} .
\end{equation}
For typical mass ratios of interest, $r_L \simeq 1/3$, and a donor
star mass-radius relation of approximately $M_d \propto R^{4/5}$ (for
CNO cycle homology), we have:
\begin{equation}
\label{esteq}
\left( \frac{\dot{M}_{\rm{wind}}}{M_{\odot}~{\rm yr}^{-1}} \right) 
\gtrsim 5 \times 10^{13} \frac{1}{{B_{\rm{s}}}^{2}} {\left( 
\frac{M_d}{M_\odot} \right)}^{-3/2} {\left( 
\frac{\dot{M}_{\rm{RLOF}}}{M_{\odot}~{\rm yr}^{-1}} \right) }^{2} ,
\end{equation}
where $B_s$ is expressed in Gauss.

This yields an estimate of the required mass-loss rate in the
magnetically coupled wind of $\dot{M}_{\rm{wind}} \sim 4 \times
10^{-10}~M_{\odot}$ yr$^{-1}$ for the illustrative parameters:
$B_s=1000$ G, $M_d=5~M_{\odot}$, and $\dot{M}_{\rm{RLOF}} =
10^{-8}~M_{\odot}$ yr$^{-1}$. Though our computational work does not
explicitly employ eq. (10), it helps illustrate the magnitude of an
irradiation-driven wind that is required to significantly affect the
evolution of binary systems.

\subsection{Irradiation-Driven Winds}

It has been suggested that a substantial stellar wind may be driven
from the donor star in a compact binary by the flux of X-radiation
that is produced by accretion onto the collapsed star
\citep{RSTE89, TL93}. Using an even smaller wind-driving efficiency than
suggested by Tavani \& London, we will show that anomalously magnetic
intermediate-mass donor stars in close binaries can be braked
sufficiently to form short-period systems.  We will also show that for
the majority of donor stars, i.e., ones with relatively weak magnetic
fields, magnetic braking via an irradiation-driven stellar wind is {\em{not}}
strong enough to drive the systems to shorter periods.  In this same
context, we note that \citet{ITY95} and \citet{ITF97} have also
suggested that an induced stellar wind could constitute an important
difference between the evolution of cataclysmic variables and low-mass
X-ray binaries.  Iben, Tutukov \& Fedorova further pointed out that a
radiation-enhanced stellar wind should have an effect on the magnetic
braking, but their calculations did not examine the consequence of
this, and they confined themselves to the orbital angular-momentum
carried away {\em directly} by the wind -- i.e. the specific orbital
angular-momentum of the donor star.

The stellar wind-loss rate is obtained by assuming that a fraction of
the accretion luminosity is converted into the kinetic energy of a
wind, such that the matter in the wind becomes marginally unbound
(consistent with our assumptions about the wind velocity). This
energy-balance argument gives:
\begin{equation}
\dot{M}_{\rm{wind}}=L_{x} f_{\Omega} f_{\epsilon} \frac{R_d}{GM_d}
\end{equation}
where the total accretion luminosity $L_x$ is multiplied by a
geometric factor $f_{\Omega}$ to find the flux that intercepts the
donor, and a wind-driving energy efficiency factor,
$f_{\epsilon}$. The X-ray luminosity is taken as some fraction, $f_x$,
of power produced by an accretion rate of ${\dot{M}}_{\rm{RLOF}}$
before the last stable orbit around a non-rotating black hole:
\begin{equation}
L_x \simeq \left(1-\frac{\sqrt{8}}{3}\right) f_x 
{{\dot{M}}_{\rm{RLOF}}} ~c^2 ~~,
\end{equation}
where c is the speed of light, $f_x$ is a rest-mass to energy
conversion factor of order unity, and the factor in brackets is the
dimensionless energy lost by the innermost stable orbit, i.e., at
three Schwarzschild radii.  Finally, we find:
\begin{equation}
\label{Mdotwind}
\dot{M}_{\rm{wind}}=\psi \frac{R_d \dot{M}_{\rm RLOF}} {GM_d} ~~,
\end{equation}
where, for convenience, we have defined:
\begin{equation}
\psi = \left(1-\frac{\sqrt{8}}{3}\right) f_x f_{\Omega} f_{\epsilon} c^2  .
\end{equation}

Initially we used a combined value for $\psi/c^{2}$ of $\sim 10^{-6}$,
based on the assumptions that the wind-driving energy conversion
efficiency $f_{\epsilon} \sim 10^{-3}$, the solid angle $f_{\Omega}
\simeq 10^{-2}$ and $f_{\rm x} \sim 1$. Clearly all these individual
values are subject to some uncertainty, but we believe that the
assumed values of $\psi$ are on the conservative side;
e.g. \citet{TL93} calculated `wind efficiencies' -- $f_{\epsilon}$ --
between $10^{-3}$ and $10^{-1}$. We began by using their lowest
calculated values for $f_{\epsilon}$, and then took $\psi$ to be one
and two orders of magnitude {\emph{lower}} (i.e., $\psi/c^{2}=10^{-7}$
and $10^{-8}$).

In the computational models we took into account the fact that the
irradiation-induced stellar wind carries away the specific
{\emph{orbital}} angular-momentum of the donor star in addition to the
braking effect it produces on the {\emph{rotation}} of the donor.  We
find that the {\em direct} loss of {\em orbital} angular-momentum via
the stellar wind is quite small, as might be expected from the fact
that the Alfv\'en radius is much greater than the dimensions of the
binary system.

\subsection[]{Analytic Results for $\dot M_{\rm RLOF}$}

A more sophisticated estimate than eq. (\ref{esteq}) for the
irradiation-induced wind loss can now be assembled. In particular, we
can form a closed set of equations, eliminate $\dot{M}_{\rm{wind}}$,
and compute the mass-transfer rate when the magnetic braking torque
just prevents the orbital separation from increasing due to the
effects of mass transfer. Beginning with eq. (\ref{MBtorque}), we
substitute for the wind-loss rate from eq. (13), and use Kepler's 3rd
law to obtain:
\begin{equation}
\label{dotJw}
\dot{J}_{\rm MB}=-B_{\rm{s}} \sqrt{ \frac{\psi \dot{M}_{\rm{RLOF}} 
M_{\rm{tot}}}{a^{3}}} \left( \frac{R_{\rm{d}}^{15}}{G M_{\rm{d}}^3} 
\right)^{1/4} ~~.
\end{equation}

The time derivative of the expression for the total angular-momentum
of the binary,
\begin{equation}
J_{\rm{sys}}=M_{\rm{d}}M_{\rm{BH}}\sqrt{\frac{G a}{M_{\rm{tot}}}} ~~~,
\end{equation}
where $M_{\rm BH}$ is the black hole mass, with the assumption of
mass conservation (i.e., $\dot{M}_{\rm{tot}}=0$) and the constraint
that $\dot{a}=0$ (i.e., we are seeking the condition where the binary
becomes neither wider nor tighter) yields:
\begin{equation}
\label{dotJs}
\dot{J}_{\rm{sys}}=\sqrt{\frac{G a}{M_{\rm{tot}}}} \left( M_{\rm{d}} 
- M_{\rm{BH}} \right) \dot{M}_{\rm{RLOF}}.
\end{equation}
Thence we can equate eqs. (\ref{dotJw}) and (\ref{dotJs}) and solve
for $\dot{M}_{\rm{RLOF}}$. We again use the parameterisation that $a =
R_s/r_L$ and define $q = M_{\rm{d}}/M_{\rm{BH}}$ to find:
\begin{equation}
\label{EqmSoln}
\dot{M}_{\rm{RLOF}}= \psi B_{\rm{s}}^{2} r_L^4 
\sqrt{\frac{R_{\rm{d}}^{7}}{G^{3}M_{\rm{d}}^{3}}} \left( 
\frac{q+1}{q-1} \right)^{2} .
\end{equation}

\begin{figure}
\begin{centering}
\epsfig{figure=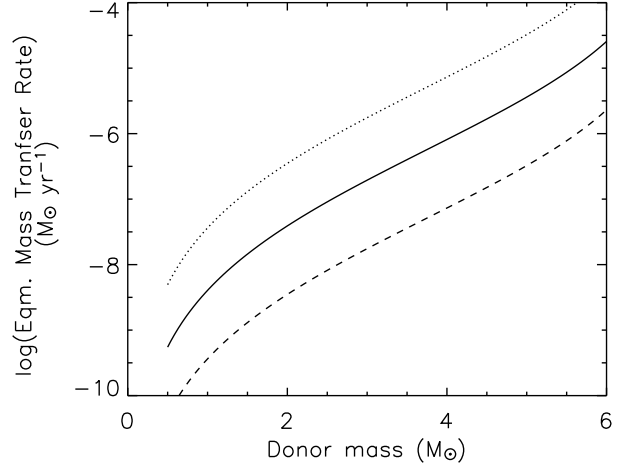, width=8cm}
\caption{\label{EqmMdot} The equilibrium mass-transfer rate solution
predicted by eq. (\ref{EqmSoln}) for a constant accretor mass of $7~M_{\odot}$.
We plot three different stellar magnetic field strengths -- 300~G (dashed),
1000~G (unbroken line), and 3000~G (dotted). Here $\psi/c^{2}=10^{-6}$.  }
\end{centering}
\end{figure}

Figure \ref{EqmMdot} illustrates the behaviour of this equation, again
assuming $R_d \propto M_d^{4/5}$. Note that the solutions scale
linearly with the overall wind-driving efficiency $\psi$, and go as
the square of $B_s$.

\subsection[]{Effects on the Canonical LMXB Population}

\begin{figure}
\begin{centering}
\epsfig{figure=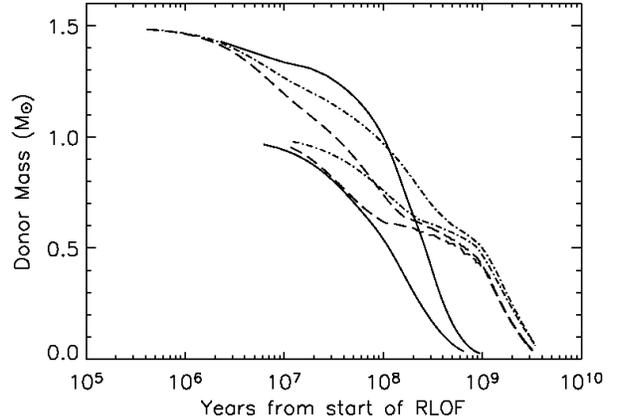, width=8cm}
\caption{\label{LMXB} We show the evolution of two LMXB systems
with initial donor masses of both 1 and 1.5 $M_{\odot}$,
where each system contains a $1.4~{\rm{M}}_{\odot}$ NS accretor.
We plot the donor mass against time; the initial orbital periods
were 8 h and 12 h, respectively. The solid lines use a standard
magnetic braking parameterisation, whilst the broken lines use
our new formalism with a constant field strength of 300 G (dot-dashed)
and 500 G (dashed). }
\end{centering}
\end{figure}

Before applying this model to Ap-star donors -- adding an
irradiation-driven wind to their intrinsically high field strengths --
we did a calculation to check that such physics would not unreasonably
distort the standard picture of low-mass X-ray binary (LMXB) evolution
during the X-ray phase. These systems contain neutron-star accretors
and low-mass donor stars, and they are typically assumed to evolve via
`conventional' magnetic braking \citep{VZ81, RVJ83} where the magnetic
field of the donor is presumably generated by dynamo action.  Stellar
dynamos are by no means precisely-solved phenomena, but following
\cite{C-CJ94} we assume that the dynamo of the donor star is saturated
at the orbital periods we consider ($\leq 1$ d). Hence we use fixed
stellar dipole field strengths of 300 and 500 G.

For this simple assumption we found a range of LMXB evolutionary
sequences that qualitatively reproduced test calculations performed
with the magnetic braking parameterisation of \cite{RVJ83}. For clarity,
the systems subject to our new mechanism do not experience conventional
magnetic braking, though we do include gravitational wave radiation.
Figure \ref{LMXB} contrasts two of these binary sequences: the
irradiation-driven braking is, for these assumptions, less strong than
conventional magnetic braking, and -- when combined with conventional
magnetic braking -- does not greatly affect the standard picture
of LMXB evolution. Note that in this case we assumed $\psi/c^{2}=10^{-6}$
-- the largest value of our wind-driving parameter.

\section[]{Black-Hole Binary Populations}

For populations of both short-period and long-period black-hole binaries to be
produced, their progenitors must somehow differ.  Something must
distinguish between those systems which are to evolve to long and to
short periods. Many of the donor stars in black-hole binary systems are
initially expected to be A and B stars (see, e.g., PRH).  We therefore
suggest that a natural distinction, within a population of intermediate-mass
stars, is between those which possess anomalously high magnetic field
strengths (i.e., the Ap and Bp stars) and those which do not 
(ordinary A/B stars).  
The magnetic braking we postulate, that takes 
place in the high-field systems 
via irradiation induced stellar 
winds, would cause a subset of the black-hole 
binaries to evolve to 
short orbital periods.

\begin{figure*}
\begin{centering}
\epsfig{figure=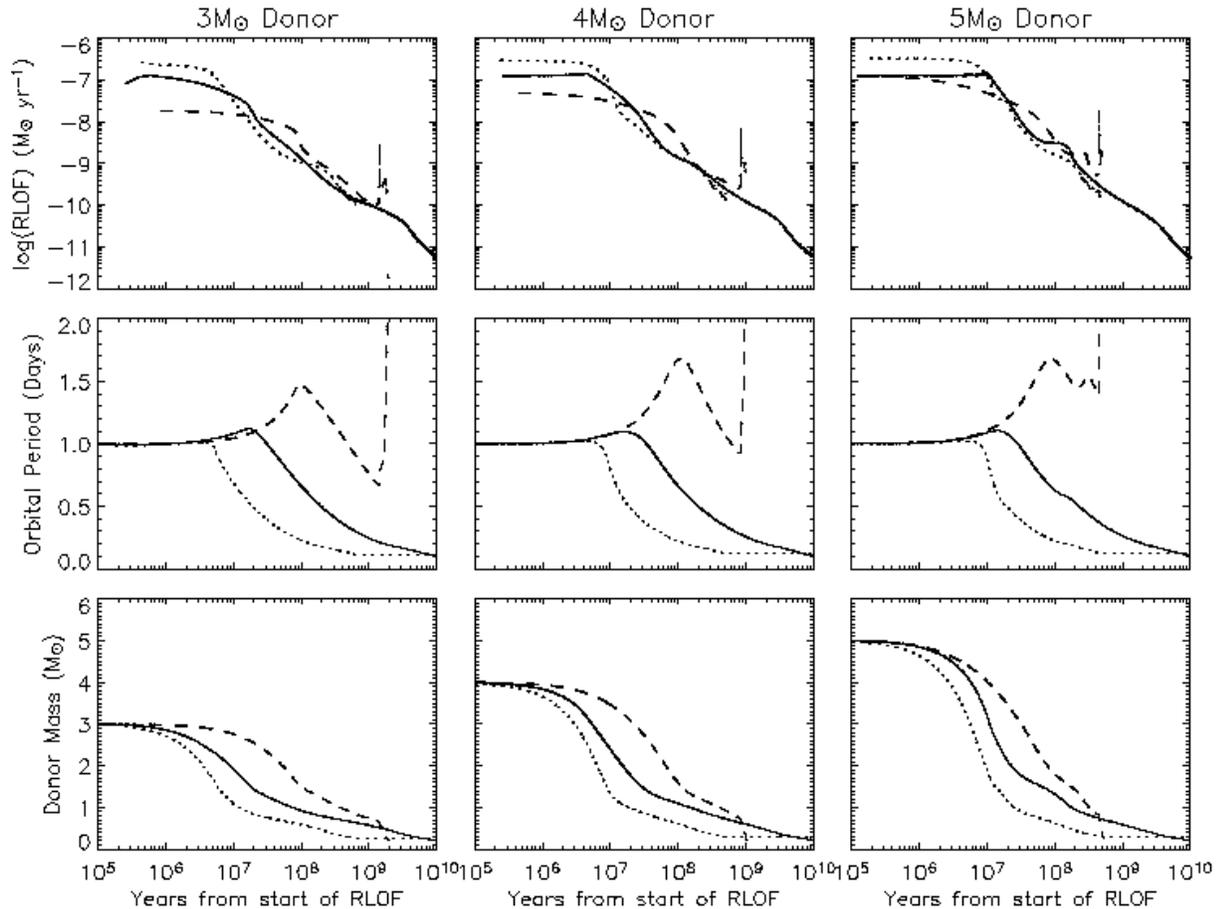, width=16cm}
\caption{\label{1E6} For a nominal wind driving energy conversion efficiency
$f_{\epsilon} = 10^{-3}$ (see text), we show the evolution of three
different companion masses (columns: left to right $3$, $4$ and
$5~{\rm{M}}_{\odot}$) for three different surface magnetic field
strengths: 300 G (dashed), 1000 G (unbroken line), and 3000 G
(dotted). For the lowest field strength, the orbital period finally
increases; also higher magnetic fields result in higher initial
mass-transfer rates, as expected (see eq. 18). The mass-transfer rates
have been slightly smoothed for clarity.  }
\end{centering}
\end{figure*}

The question of what happens to Ap star field strengths when
Roche-lobe overflow begins is not clear, and in this work we assume
that the initially strong field persists. If the flux is a fossil
field frozen into the stellar matter \citep[e.g.][]{BS04}, the
anomalous field may be lost with the mass in the envelope as it is
transferred to the black hole. Note, however, that the field has
to persist only long enough for a dynamo to begin operating as the
stellar mass decreases. There has also been a suggestion by
\citet{TWF04} that the magnetic fields seen in white dwarfs may be
directly linked to the magnetic flux of the progenitor star
(specifically with reference to Ap and Bp stars), in which case the
fields would have had to survive evolutionary mass loss, if not
necessarily Roche-lobe overflow. This suggestion is tenuous, and the
fate of the stellar magnetic field during RLOF is a significant
unknown in this picture -- not just with respect to the initial flux,
but also the possible operation of a dynamo. We do not include any
stellar dynamo in the current calculation; rather we simply assume
that $B_s$ is a constant over time, even during mass loss (in \S 3.2 we
discuss the consequences of non-constant surface field strengths).

We use an updated version of Eggleton's stellar evolution code
\citep[e.g.][]{PPE71, PTEH95} to evolve systems with initial donor
stars of $3$, $4$, and $5~{\rm{M}}_{\odot}$ orbiting a black hole of
mass $7~{\rm{M}}_{\odot}$ with an initial orbital period of one
day. We use the RLOF rate from the previous time-step to calculate the
magnitude of the irradiation-driven stellar wind (eq. \ref{Mdotwind}),
and from this the magnetospheric radius (eq. \ref{MagRad}), and
finally the angular-momentum loss (eq. \ref{MBtorque}) in the current
timestep. Overall, this is equivalent to using eq. (\ref{dotJw}).

Figure \ref{1E6} shows the outcomes of these evolutionary sequences
for a range of stellar magnetic field strengths, and with a
wind-driving parameter $\psi/c^{2}=10^{-6}$.  With these parameters,
only systems with donor stars having $B_s \lesssim 300$ G will evolve
to wide orbits; for larger dipole $B$ fields the systems will become
compact during the mass-transfer phase.

In Fig. \ref{1E6}, all the systems begin RLOF with a high
mass-transfer-rate plateau lasting $\sim 1\,$--\,100\,Myr (only as long as
100\,Myr for the 300\,G stellar fields), at the end of which a maximum
in the orbital period evolution is reached. Donor stars with only 300\,G fields
reach core hydrogen exhaustion after $\sim 1$\,Gyr, and their
orbital periods lengthen as they expand. The donor stars with stronger
fields -- having lost mass more rapidly -- evolve more slowly,
continue burning hydrogen in their cores out to $> 10\,$Gyr, and are
good candidates for short-period black-hole binaries. They have all
dropped below 1 ${\rm{M}}_{\odot}$ by $\sim 100$\,Myr after the onset of
RLOF; likewise, they spend relatively little time above an orbital
period of $\sim$0.5 days, which agrees well with the orbital
parameters in table 1 of \citet{LBW}. At 1\,Gyr, with periods of
$\sim 0.25\,$d and donor masses around 0.5\,${\rm{M}}_{\odot}$, the
1\,kG models produce mass transfer rates of
$\sim10^{-10}~{\rm{M}}_{\odot} {\rm{yr}}^{-1}$.

We note explicitly that these systems spend most of their evolution
with low-mass donors and are most likely to be observed as low-mass
systems.

We have also calculated a range of binary sequences for less efficient
wind-driving energetics, specifically $\psi / c^{2}=10^{-7}$ and
$10^{-8}$.  For these lower wind-driving efficiency factors, higher
magnetic fields for the donor star are required to obtain results
similar to those shown in Fig. \ref{1E6} (see eq. 18). In particular,
$B_s$ must scale as $\psi^{-1/2}$.

These sequences demonstrate that the anomalous magnetic braking
scenario is able to produce short-period systems for reasonable
magnetic field strengths and conservative assumptions about the
irradiation driven stellar winds.

\subsection[]{The Long and the Short: Population Statistics}

In order for the proposed anomalous magnetic braking scenario to be
viable for producing short period black-hole binaries, it should yield
the correct ratio of long to short period systems.  From the small
proportion of A-stars that have anomalously high magnetic fields
\citep[around 5\% -- see, e.g.][]{Land82, SWDetc02}, we might expect
the population of short-period black-hole binaries to be much smaller than
that of long-period systems, whilst the observed populations are
roughly equal in size. However, the systems driven to lower masses
more quickly by magnetic braking will live longer than those which are
not, and so an imbalance in the population birthrates (of around a
factor of 20) is counteracted by the longer lifetimes of the
short-period systems.

\begin{figure}
\begin{centering}
\epsfig{figure=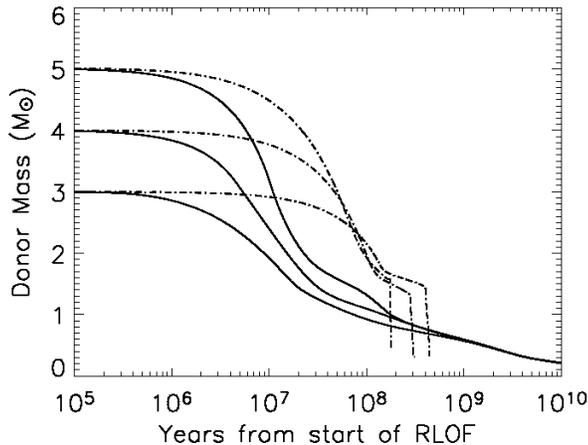, width=7.7cm}
\caption{\label{lifetimes} In order to illustrate the contrast in
lifetimes for magnetic and non-magnetic donor stars, we compare the
mass evolution of systems with magnetic fields of 1000 G (solid lines,
as in Fig. \ref{1E6}) and $B_s = 0$ (broken lines) for initial donor
masses of $3$, $4$, and $5M_{\odot}$ and an orbital period at contact
of 1 d. Nuclear evolution and the complete transfer of the stellar
envelope terminate the systems without magnetic braking (which are
evolving to longer periods) after only a few hundred Myr. The
short-period systems are still transferring mass after 10 Gyr (for the
corresponding mass transfer rates, see Fig. \ref{1E6}).  }
\end{centering}
\end{figure}

A representative lifetime estimate for the short-period systems is several
Gyr (the 1000 G curves in Fig. \ref{1E6} all continue indicating mass
transfer until 10 Gyr, so the lifetime depends on choosing a minimum
mass-transfer rate cutoff). Systems with the same initial orbital
period, but with zero magnetic field, live only 200--400 Myr (see
Fig. \ref{lifetimes}). This ratio of lifetimes -- demonstrated
explicitly in Fig. \ref{lifetimes} -- is approximately the required factor
of 20 (to compensate for the small fraction of Ap stars), and shows that 
our model is naturally consistent with the observed numbers. 

We note that relaxing our simplifying assumption -- that the magnetic
field strength does not decrease as the donor is stripped of mass --
should make the lifetimes of the short-period systems even longer, as
in the present calculations the low-mass donor stars are still subject
to significant magnetic braking. More detailed lifetime arguments are
also likely to depend on how evolved the donors are at contact and hence
on the post-common-envelope orbital periods.

Wary of these unknown details, we calculated a range of simple population
distributions to examine the feasibility of matching the observations of
these low-mass black-hole binaries. A detailed presentation of these results
is not included but we can draw an important general conclusion. The period
distribution does seem to be best matched by decreasing the surface magnetic
field strength when the donor mass falls below $\sim 1~M_{\odot}$; such
evolution of the initially strong Ap field is eminently reasonable.
(We repeat, however, that for all of the results shown in Figs. 2--5,
the magnetic field strength was held constant.)

We note that -- though we expect the stellar surface field strength 
to decay -- the onset of transient behaviour would naturally reduce 
the efficiency of this irradiation-driven braking mechanism. 
Equation (\ref{MBtorque}) reveals that the torque is proportional 
to the square root of the wind mass loss, and hence 
(via eq. \ref{Mdotwind}) to the square root of the mass 
accretion rate. Persistent accretion is thus more efficient at 
driving braking than phases of outburst and quiescence, which 
is an effect that should be considered in more detailed 
future investigations of this mechanism. 
We also acknowledge that, in systems where the accretion rate 
begins to exceed the Eddington limit, the irradiation 
efficiency (and hence $\psi$) will drop; however, since the mass 
transfer is itself driven by our braking mechanism, this effect is to some 
extent self-regulating and would not necessarily result in 
widening of the binary.

\subsection{Transient Behaviour}

We checked the consistency of our models for compact
black-hole binaries for their potential behaviour as soft-X-ray
transients via the familiar thermal-ionization disk instability
\citep{CGW82, KKB96, vP96, DHL01, L01}.  At each step in the binary
evolution we calculated the susceptibility of the accretion disc to
instability, either by computing the temperature that would be attained
in the outer portions of the accretion disk due to X-ray irradiation
\citep[using eq. (8) in][]{RPP05}, or by determining a critical accretion
rate using expressions due to \citet{KKS97} and \citet{DLHC99}. Though
these different criteria yield considerable freedom of outcome, we find that
all the methods for determining transient behaviour can match the observations: they 
agree that systems with $P_{\rm orb} \lesssim 0.5$ days and $M_{\rm d} 
\lesssim 1 M_{\odot}$ are transient (provided the magnetic braking at 
that epoch is not implausibly strong). 

Common to all the criteria for transient behaviour is the prediction 
that during the
initial phase of mass transfer -- as the donor approaches $\sim 1 M_{\odot}$
-- the systems will be persistent sources. These systems will be evolving
rapidly and thus there may be too few of them to have been discovered.
However, the model predicts that there is a class of black-hole binaries
with intermediate-mass donors and orbital periods $\lesssim 1$ day that
are steady X-ray emitters.

In light of the fact that the wind-driving mechanism becomes 
significantly less efficient in transient systems, we point out that 
at the periods where the systems are predicted to become transient 
the effective temperatures of the donors have dropped within the 
range where `normal' magnetic braking operates, i.e. the stars are 
capable of driving their own winds.

\subsection[]{Observational Test: CNO Abundances}

\begin{figure}
\begin{centering}
\epsfig{figure=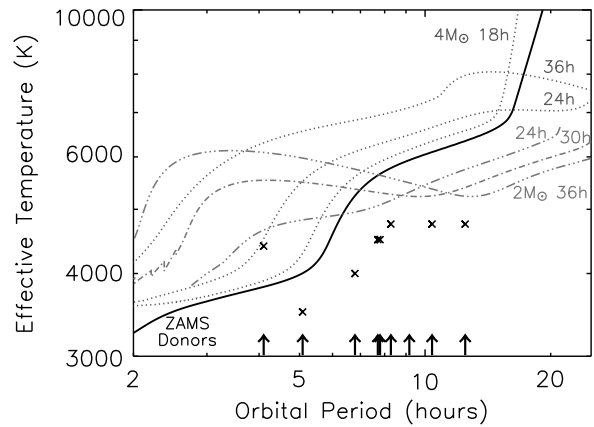, width=8.5cm}
\caption{\label{TeffPplane}
Evolutionary tracks in the donor-mass--period plane 
for donors initially of $4~M_{\odot}$ with initial periods of 18, 24 
and 36 h (dotted lines), as well as for donors initially of 
$2~M_{\odot}$ with initial periods of 24, 30 and 36 h (dash-dotted 
lines). We also indicate the periods of known compact black-hole 
X-ray binaries (arrows on period axis), with crosses marking the 
{\emph{earliest}} spectral types given in \protect\cite{C05}. The solid line 
marks zero-age main sequence (ZAMS) donors which would fill their 
Roche lobes with a $7~M_{\odot}$ companion. Note that less evolved 
donors are closest to the ZAMS line, and for the $4~M_{\odot}$ donor 
more evolved donors lead to earlier spectral types, whilst more 
evolved $2~M_{\odot}$ stars are cooler at longer orbital periods but 
become hotter as the interior is further exposed.}
\end{centering}
\end{figure}

As well as population arguments, we can use individual objects to 
test the proposed
anomalous magnetic braking mechanism. Observations that show 
CNO-processed elements
on the surface of XTE J1118+480 (Haswell et al. 2002; 
the system has a 4.1 hour orbital period)
are strong evidence that the donor is the descendent of an intermediate-mass
star. Under the assumption that only dynamo-driven magnetic braking 
occurs, Haswell
et al. (2002) found that only a very small parameter space could have produced
XTE J1118+480 with such a CNO-processing signature, with a donor star of initial
mass $1.5~{\rm{M}}_{\odot}$. Our irradiation wind-driven magnetic braking model
relaxes these constraints considerably.

Moreover, the far UV spectrum taken by Haswell et al. appears to lack 
emission in
oxygen as well as carbon. This fact is explained most naturally if 
the progenitor
was more massive than $1.5~{\rm{M}}_{\odot}$, such that the full CNO 
tricycle could
have operated. As PRH suggested, progenitors with such large masses would
{\emph{require}} a new angular-momentum loss mechanism -- such as the 
one proposed
in this work.

\subsection{Observational Test: Spectral Types}

Figure \ref{TeffPplane} compares the results of our binary evolution
calculations with unevolved, primordially low-mass donors in the
$T_{\rm eff}-P_{\rm orb}$ plane, and marks the periods of the known 
compact black-hole X-ray binary systems. The companion stars in 
these latter binaries are observed to have cool 
spectral types \citep[see, e.g. ][where the donors are listed 
as broadly mid-K type stars; \citet{Tetal04} discuss their 
spectral type determination for XTE J1118+480]{C05}. 
This forms a strong constraint on intermediate-mass
progenitor stars: donors which have been allowed to evolve by the onset
of mass transfer tend to have effective temperatures higher
than required at the known orbital periods (see Fig. \ref{TeffPplane}). 
Note that the behaviour of the $4~{\rm{M}}_{\odot}$ donor in 
Fig. \ref{TeffPplane} contrasts strongly with that predicted for the 
donors in CVs, where more evolved donors produce later spectral types 
\citep[see, e.g.][]{BBKW98, BK00}.

We stress that the crosses in Fig. \ref{TeffPplane} mark the
{\emph{hottest}} temperatures consistent with the observed spectral types.
This difference between theory and observation is hard to reconcile even
when approximate non-grey atmospheric corrections
are applied to our models \citep[see][]{CB97, Betal98, PHR03}. 
At best, we can force the models to approach the temperatures of 
primordially 
low-mass ZAMS stars by having the donors begin 
transferring mass to the black 
hole early on the main sequence. 
However, even these values of $T_{\rm eff}$ 
are still too high.

The conversion between spectral type and effective temperature is 
non-trivial and it may be possible for our coolest models -- 
late-G type donors  -- to be mistaken for mid-K stars, 
but we consider it unlikely that such a conversion error 
alone could account for the wide discrepancy between 
the bulk of our models and the observations.  
However we note that \citet{Tetal04} find that 
the donor star contributed only $\sim55\%$ of the light in their 
observations of XTE J1118+480 during quiescence and suggest that 
unambiguous determination of donor spectral types in such systems is 
challenging. 

We did consider one potential mechanism by which the secondaries 
in these systems might
appear cooler than our models predict. The companion in the black-hole 
binary GRO J1655-40 (Nova Scorpii 1994, V1033 Sco) is known to be 
polluted with several $\alpha$-process elements
\citep[for a detailed investigation, see][]{PNMNMS}; such pollution
would be expected to increase the opacity of the affected layers.
We investigated both ad-hoc opacity modifications for the surface layers and
uniform composition changes. Although we managed to produce closer
correspondence with the observations, we cannot claim that the 
improvements were very dramatic.

\section[]{Summary and Conclusions}

In this paper we have proposed that the subset of neutron-star
and black-hole binaries that form with Ap or Bp star companions
will experience systemic angular-momentum losses due to magnetic
braking, not otherwise operative with intermediate-mass companion
stars.  We have quantified how a magnetically coupled,
irradiation-driven stellar wind can lead to substantial loss of
systemic angular-momentum.  We have demonstrated with detailed
binary stellar evolution calculations that the proposed magnetic
braking scenario involving Ap/Bp donor stars is effective in allowing
such systems to evolve to short orbital periods ($P_{\rm orb}
\lesssim 10$ hr).  In the absence of such magnetic braking, binaries
where intermediate-mass donor stars transfer mass onto more massive
black holes inevitably evolve to long orbital periods,
i.e., $P_{\rm orb} \gtrsim$ weeks.

Our Ap/Bp magnetic braking scenario has been applied to a
specific astrophysical problem, namely the formation and evolution of compact
black-hole binaries with low-mass companion stars. A number of previous studies
have encountered difficulties with explaining the formation of such systems
(PRH and references within).  This led PRH to consider some form of anomalous
magnetic braking to prevent systems from evolving to long orbital 
periods.  In this
work, we have presented such a mechanism that requires no new physics 
-- and, indeed,
is a consequence of previously published mechanisms acting in close binaries.

One of the conceptual difficulties with the formation of short-period
black-hole binaries is that a low-mass donor star seems incapable of
ejecting the common envelope during the prior formation of the black hole
(PRH and references within; see also the Appendix).  More massive stars, including stars of
intermediate mass (i.e., $3-5~M_\odot$), would be more likely to be able to
eject the common envelope.  Even under favourable (or optimistic) conditions
in which a low-mass donor might conceivably be able to eject the
common-envelope of the black-hole progenitor, stellar demographics suggest
that the formation of binary systems with mass ratios of around 20 is 
very uncommon \citep[see, e.g.][]{GCM80}.

On the other hand, intermediate-mass stars are not thought to be subject to
magnetic braking.  As mentioned above, such angular-momentum losses 
are required
to prevent the system from evolving to longer, rather than shorter, 
orbital periods. 
We have applied our Ap/Bp magnetic braking scenario to this problem in order
to solve the difficulties with both the common-envelope ejection and the
angular-momentum loss mechanism.

We have calculated a sequence of binary evolution models that demonstrate the
Ap/Bp magnetic braking model is successful at reproducing the short orbital
periods and low donor masses observed for the compact black-hole binaries. 
Moreover, we have shown that, since these systems have a longer lifetime
than the corresponding systems without magnetic braking, the relative
populations of compact and wide black-hole binaries could be explained
in spite of the small fraction of A/B stars that are of the Ap/Bp class.
Our model also helps to explain the evidence for CNO-processed material
seen at the surface of XTE J1118+480.

As demonstrated in Fig. 5, however, the effective temperatures of our
model donor stars are significantly higher than for those of the observed
donor stars. This seems to be a generic difficulty with any formation scenario
that invokes primordially intermediate-mass donor stars, and is not 
specifically related to our suggested angular-momentum loss 
mechanism. Any new angular-momentum loss mechanism proposed to 
explain the formation of short-period black-hole binaries would have 
to account for the same mismatch in spectral types. 
Hence in developing our model, we may thus have established a
substantial constraint on the broad set of formation scenarios which begin with
intermediate-mass secondaries.

The problem results from the fact that even though the initially
intermediate-mass star loses much of its mass fairly rapidly (driven by
magnetic braking), the star is still somewhat evolved chemically (both
in He and CNO abundance) by the time its mass has been reduced 
to $\sim1~M_\odot$. 
This prevents the donor stars from achieving the cooler
observed values of $T_{\rm eff}$.  Even in the limit that 
the intermediate-mass star could
lose much of its mass within a few of its thermal timescales, the
subsequent evolution would not automatically lead to the evolutionary state of
the observed donor stars, i.e., undermassive and underluminous for a given
orbital period. 

Even though the systems that we have generated have donor stars with 
hotter effective temperatures than are observed, our 
analysis predicts that such systems should exist and could be 
discovered in the future.  The signature would be orbital periods 
between $\sim5-15$ hours,
donor-star masses of $\sim 0.5-1~M_\odot$, and effective temperatures
$5000-7000$ K.  If black-hole binaries with these higher effective temperatures
are not found, then this would imply that either the donor stars are
rarely, if ever, of the Ap/Bp variety, that the proposed magnetic braking
mechanism is not as efficient as we have calculated, or that the Ap/Bp magnetic
fields do not persist as mass is lost from the star.

In a subsequent paper we plan to explore further scenarios for the formation
and evolution of compact black-hole binaries.  These include (i) rapid
mass loss of the intermediate-mass secondary at the end of the common
envelope phase; (ii) explosive common-envelope ejection \citep[see, 
e.g.][]{IP03};
(iii) formation of the low-mass donors in situ from the remnants
of a failed common-envelope ejection; and (iv) evolutionary paths of 
low-mass 
donors that lead to the system properties that are 
observed.  

\section*{Acknowledgements}
We thank Henk Spruit for very stimulating discussions at the 
inception of this work. We also thank Ron Remillard for valuable input 
regarding the
properties of black-hole binaries. One of us (SR) acknowledges support from 
NASA contract TM5-6003X, whilst SJ is supported by PPARC grant 
PPA/G/S/2003/00056.

\appendix
\section*{Appendix: Common-Envelope Energetics}
Herein we briefly summarise the arguments that imply that a low-mass
donor star cannot eject the envelope of a BH-producing star.
In the standard formulation using simple energetic arguments, the
expression for the final orbital separation at the end of the
common-envelope phase, $a_f$, to that just at the onset of mass transfer
from the primary to the secondary, $a_i$, is, to a good approximation:
\begin{equation}
\label{CEsepar}
\frac{a_f}{a_i} \simeq \frac{r_L}{2} \left(\frac{M_c}{M_1M_e}\right) 
M_2 \lambda ~~~, \tag{A1}
\end{equation}
(e.g., Webbink 1985; Dewi \& Tauris 2000; Pfahl, Rappaport, \& 
Podsiadlowski 2003;
Rappaport, Podsiadlowski, \& Pfahl 2005) where $M_1$, $M_c$, and $M_e$ are the
masses of the black-hole progenitor, its core, and its envelope, respectively,
$M_2$ is the secondary mass, and $r_L$ is the size of the progenitor's Roche
lobe radius in units of the orbital separation.  In this expression, $\lambda$
is a parameter that gives the binding energy, $E_{\rm bind}$, of the 
envelope of
the primary in the expression $E_{\rm bind} \equiv GM_1M_e/(R_2 
\lambda)$, where
$R_2$ is the radius of the primary.  We have assumed here that the 
efficiency with
which the gravitational potential energy between the core of the 
primary and the
secondary star can be used to eject the primary's envelope is unity. 
For values
less than unity, the following argument is only strengthened.

The quantity in parentheses on the right hand side of eq. 
(\ref{CEsepar}) has a numerical
value of 0.022 $\pm 0.002~M_\odot^{-1}$ for all primary masses 
between 25 and 45
$M_\odot$ which we consider to be the main source of the black holes 
in question.
For larger primary masses, it only becomes more difficult to remove 
the envelope.
We can then write eq. (\ref{CEsepar}) as:
\begin{equation}
a_f \simeq 0.011 (a_i r_L) M_2 \lambda~~~, \tag{A2}
\end{equation}
where the quantity $a_i r_L$ is just the radius of the donor star at 
the onset of
the common envelope phase.  For typical progenitor masses of the 
black holes, we
take the maximum stellar radius to be $\sim$2300 $R_\odot$.  Thus, at 
the end of
the common-envelope phase the maximum orbital separation is given by:
\begin{equation}
a_f \simeq 25 \left(\frac{M_2}{M_\odot}\right)
\left(\frac{R_{\rm max}}{2300 \, R_\odot}\right) \lambda ~~ R_\odot ~~~. \tag{A3}
\end{equation}

In order to avoid a merger of the black-hole progenitor core and the 
secondary at
the end of the common-envelope phase, the Roche lobe of the secondary must be
larger than a main-sequence star of the secondary's mass.  If we use a simple
mass-radius relation for low-mass main sequence stars, i.e., $R_2 = 0.85
(M_2/M_\odot)^{0.85} R_\odot$, and employ the Roche-lobe radius expression for
low-mass stars due to Paczy\'nski (1971), we find the following lower limit for
the parameter $\lambda$ in order to avoid a merger of the core and secondary:
\begin{equation}
\label{LambdaIneq}
\lambda \ga 0.15 \left(\frac{2300 \, R_\odot}{R_{\rm max}}\right)
\frac{1}{\sqrt{M_2}}~~~. \tag{A4}
\end{equation}
The numerical coefficient in eq. (\ref{LambdaIneq}) is good to $\pm$15\% over
a wide range of primary and secondary masses.

Thus, for initially low-mass secondaries, i.e., $M_2  \lesssim 1 \, M_\odot$
the $\lambda$  parameter must be $\ga 0.15$ even when the common 
envelope occurs
at the widest plausible binary separation.  For smaller orbital 
separations at the
onset of the common-envelope, the values of $\lambda$ required to 
avoid a merger
approach unity. Figure 1 of PRH illustrates this binding energy parameter for
stars capable of forming a BH (see also Dewi \& Tauris 2000). 
For the calculations and assumptions made by PRH
-- including wind mass loss according to \cite{NdJ} -- stars of 25 $M_{\odot}$
and greater have total binding energies such that
$\lambda < 0.1$ for all $R > 300 R_{\odot}$. Note that those values 
for $\lambda$
include thermal terms, and hence assume that all the energy of 
recombination for
the ionised species in the envelope can be liberated to aid the common-envelope
ejection process.

\label{lastpage}
\end{document}